\documentclass{aa}  

\usepackage{graphicx}
\usepackage{txfonts}
\begin{document}

   \title{First dust measurements with the Solar Orbiter Radio and Plasma Wave instrument}

   \author{A.~Zaslavsky \inst{1}
          \and
          I.~Mann \inst{2}
          \and
          J.~Soucek \inst{3}
          \and
          A.~Czechowski \inst{4}
          \and
          D.~Pisa \inst{3}
          \and
          J.~Vaverka \inst{5}
          \and
          N.~Meyer-Vernet \inst{1}
          \and
          M.~Maksimovic\inst{1}
          \and
          E.~Lorfèvre \inst{6}         
          \and
          K.~Issautier \inst{1}
          \and
          K.~Racković Babić\inst{1, 7}
          \and          
          S.~D.~Bale \inst{8,9}             
          \and
          M.~Morooka \inst{10}          
          \and
          A.~Vecchio\inst{11}
          \and
          T.~Chust\inst{12}
          \and
          Y.~Khotyaintsev\inst{10}
          \and
          V.~Krasnoselskikh\inst{13}
          \and
          M.~Kretzschmar\inst{13, 14}
          \and
          D.~Plettemeier \inst{15}
          \and
          M.~Steller\inst{16}
          \and
          Š.~Štverák\inst{3, 17}
          \and
          P.~Trávníček\inst{8, 17}
          \and
          A.~Vaivads\inst{10, 18}
         }

   \institute{LESIA, Observatoire de Paris, Université PSL, CNRS, Sorbonne Université, Université de Paris, France\\
              \email{arnaud.zaslavsky@obspm.fr}
         \and
         Arctic University of Norway, Tromsø, Norway
         \and
         Institute of Atmospheric Physics of the Czech Academy of Sciences, Prague, Czechia
         \and
         Space Research Center, Polish Academy of Sciences, Warsaw, Poland
         \and
         Faculty of Mathematics and Physics, Charles University, Prague, Czech Republic
         \and
	CNES, 18 Avenue Edouard Belin, 31400 Toulouse, France
	\and
	Department of astronomy, Faculty of Mathematics, University of Belgrade, Serbia
	\and
	Space Sciences Laboratory, University of California, Berkeley, CA, USA
	\and
	Physics Department, University of California, Berkeley, CA, USA
	\and
	Swedish Institute of Space Physics, Uppsala, Sweden  	
	\and
	Radboud Radio Lab, Department of Astrophysics, Radboud University, Nijmegen, The Netherlands
	\and
	LPP, CNRS, Ecole Polytechnique, Sorbonne Universit\'{e}, Observatoire de Paris, Universit\'{e} Paris-Saclay, Palaiseau, Paris, France
	\and
	LPC2E, CNRS, 3A avenue de la Recherche Scientifique, Orléans, France  
	\and
	Université d'Orléans, Orléans, France 
	\and
	Technische Universität Dresden, Würzburger Str. 35, D-01187 Dresden, Germany
	\and
	Space Research Institute, Austrian Academy of Sciences, Graz, Austria
	\and
	Astronomical Institute of the Czech Academy of Sciences, Prague, Czechia
	\and
	Department of Space and Plasma Physics, School of Electrical Engineering and Computer Science, Royal Institute of Technology, Stockholm, Sweden
	}

 
 \abstract
   {Impacts of dust grains on spacecraft are known to produce typical impulsive signals in the voltage waveform recorded at the terminals of electric antennas. Such signals are, as could be expected, routinely detected by the Time Domain Sampler (TDS) system of the Radio and Plasma Waves (RPW) instrument aboard Solar Orbiter. }
  {We investigate the capabilities of RPW in terms of interplanetary dust studies and present the first analysis of dust impacts recorded by this instrument. Our purpose is to characterize the dust population observed in terms of size, flux and velocity.}
   {We briefly discuss previously developed models of voltage pulses generation after a dust impact onto a spacecraft and present the relevant technical parameters for Solar Orbiter RPW as a dust detector. Then we present the statistical analysis of the dust impacts recorded by RPW/TDS  from April 20th, 2020 to February 27th, 2021 between 0.5 AU and 1 AU.}
   {The study of the dust impact rate along Solar Orbiter's orbit shows that the dust population studied presents a radial velocity component directed outward from the Sun, the order of magnitude of which can be roughly estimated as $v_{r, dust} \simeq 50$ km.$s^{-1}$. This is consistent with the flux of impactors being dominated by $\beta$-meteoroids. We estimate the cumulative flux of these grains at 1 AU to be roughly $F_\beta \simeq 8\times 10^{-5} $ m$^{-2}$s$^{-1}$, for particles of radius $r \gtrsim 100$ nm. The power law index $\delta$ of the cumulative mass flux of the impactors is evaluated by two differents methods (direct observations of voltage pulses and indirect effect on the impact rate dependency on the impact speed). Both methods give a result $\delta \simeq 0.3-0.4$.}
   {Solar Orbiter RPW proves to be a suitable instrument for interplanetary dust studies, and the dust detection algorithm implemented in the TDS subsystem an efficient tool for fluxes estimation. These first results are promising for the continuation of the mission, in particular for the in-situ study of the dust cloud outside the ecliptic plane, which Solar Orbiter will be the first spacecraft to explore.}
   

   \maketitle
%

\section{Introduction}

For several decades, radio and plasma wave instruments have demonstrated their ability to probe dust in different space environments. Voyager's plasma wave instrument \citep{Gurnett_etal_1983} and planetary radio astronomy experiment \citep{Aubier_etal_1983} both observed broadband signals interpreted as produced by dust impacts during the crossing of Saturn’s E ring by Voyager 1 and G ring by Voyager 2 -- the plasma wave instrument, operated in dipole mode with a roughly symmetrical configuration, observing smaller amplitude signals that the radio experiment, operated in monopole mode. The technique was used again along Voyager 2 orbit, with dust measurements at Uranus \citep{NMV_etal_1986, Gurnett_etal_1987} and Neptune \citep{Gurnett_etal_JGR1991, Pedersen_etal_1991}. Voyager measurements at the outer solar system's planets were followed by others in space environments with an expected high dust flux, like cometary trails with e.g. VEGA 2's plasma wave instrument at comet Halley \citep{Oberc_1990}.

From the years 2000's or so, radio analyzers aboard missions such as Wind \citep{Wind_Waves_95},  Cassini \citep{Gurnett_RPWS_2004} or STEREO \citep{Bougeret_SWAVES_SSR08} recorded a large number of electric waveforms characteristic of dust impacts. The improvement in the technical performance of these radio detectors compared to the previous generation (the higher sampling frequency of the waveform analyzers in particular) and the large number of examples available to study has led to a better understanding of the mechanisms involved in the voltage pulses generation after a dust impact. Several recent studies detail this work of modelling and comparison to available data, such as works by \cite{Zaslavsky_2015} on STEREO, \cite{MeyerVernet_etal_2017} or \cite{Ye_etal_RPWS_2019} on Cassini and \cite{Vaverka_etal_2019} on MMS. The paper by \cite{Mann_etal_2019} provides a complete summary of the works performed on various spacecraft, a prospect for missions Parker Solar Probe and Solar Orbiter and a review of the voltage pulse mechanism in our present state of understanding. The paper by \cite{Lhotka_etal_MMS2020} also presents a detailed analysis of spacecraft charging processes in various plasma environments and an application to dust impacts on MMS. 

To quickly summarize these models, voltage pulses result from the production of free electric charges by impact ionization after a grain of solid matter hits the spacecraft. These charges modify the spacecraft (and, depending on the impact location, the antennas) potential through two main effects, one being the perturbation of the electric current equilibrium between the spacecraft and the surrounding plasma due to the collection by the spacecraft of some of these free electric charges, the other being the perturbation of the spacecraft potential by electrostatic influence from these free charges that occurs when the impact ions/electrons cloud is not neutral on overall (which happens after some of the charges have been collected or have escaped away from the spacecraft).

Along the years, and thanks to these refinements in the pulses modeling, radio analyzers have thus proven to be able to provide robust estimates of dust fluxes in various mass ranges, varying from the nanometer to the micron. Examples of successful use of this technique to derive dust fluxes include the detection of nanometer sized dust with STEREO/WAVES \citep{MeyerVernet_etal_SolPhys_Nano_2009} and Cassini/RPWS \citep{MeyerVernet_etal_Cassini_GRL_2009}, measurements of the interstellar dust flux and direction at 1 AU by STEREO/WAVES \citep{Zaslavsky_etal_2012} and Wind/WAVES \citep{Malaspina_etal_2014} or measurements of the micron to ten microns sized dust density in the vicinity of Saturn by Cassini/RPWS \citep{Ye_etal_2014, Ye_etal_2018}.

The present paper, in the continuation of these works, is devoted to the study of the dust impact data recorded by the Radio and Plasma Waves instrument, and to the derivation of the interplanetary dust fluxes along Solar Orbiter's orbit. This is of particular importance since in-situ measurements of interplanetary dust in the inner heliosphere, which are necessary to constrain and validate dust production models, are not numerous. Notably there are only few data on the dust collision fragments that form in the inner solar system and then are ejected outward. Flux estimates for these dust grains, denoted as $\beta$-meteoroids, were made based on Helios observations \citep{Zook_Berg_1975} and based on Ulysses observations \citep{Wehry_Mann_1999}. The dust collision evolution inside 1AU were studied with model calculations having only few observational constrains \citep{Mann_etal_SSR2004}. 

Recently the Parker Solar Probe (PSP) operates the FIELDS instrument \citep{Bale_FIELDS_2016} which provides observations of dust impacts. The PSP dust observations have been presented in a number of recent works \citep{PSP_Page_2020, PSP_Szalay_2020, PSP_Malaspina_2020}. These observations provide a number of interesting results on dust fluxes in the close vicinity of the Sun. Interesting in the context of this paper are the works on the second orbit of PSP with dust measurements between ca. 0.16 and 0.6 AU. \cite{PSP_Szalay_2020} showed that the observations during the second solar encounter could be explained with particles that form as collision fragments near the Sun and then are ejected by the radiation pressure force. \cite{PSP_Mann_Czechowski_2021} showed that the same fluxes could be explained with a model that combines collisional production of dust particles and their dynamics influenced by gravity, radiation pressure and Lorentz force; the latter was found to have only a small effect on the particles that were observed with PSP during the second orbit. 

In this paper we shall first present and discuss, in section \ref{sec-dust-rpw}, the waveform data recorded by the instrument and the specificities of Solar Orbiter as a dust detector. Then in section \ref{sec-stat-analysis} we focus on the statistical study of the time repartition and of the voltage amplitudes of the hits recorded. Finally, in a last section, we build on this statistical study to determine the flux of the dust population observed, and compare our results to the one obtained by other missions and to theoretical predictions.

\section{Dust measurements with RPW}
\label{sec-dust-rpw}

The RPW instrument aboard Solar Orbiter, a complete description of which is given by \cite{Solo_RPW_2020}, is designed to measure and analyze the electric field fluctuations from near-DC to 16 MHz and magnetic fluctuations from several Hz to 0.5 MHz. In this article we are mainly interested by the electric waveforms provided by the time domain sampler (TDS) subsystem of RPW. TDS provides digitized snapshots of the voltage measured between the terminals of two of the spacecraft electric antennas (dipole mode) or between one the spacecraft electric antennas and the spacecraft ground (monopole mode). The waveforms used in this article were sampled at a rate of 262.1 kHz, after being processed by an analog high-pass filter with a cutoff around 100 Hz.

\begin{figure*}
   \centering
   \includegraphics[scale = 0.4]{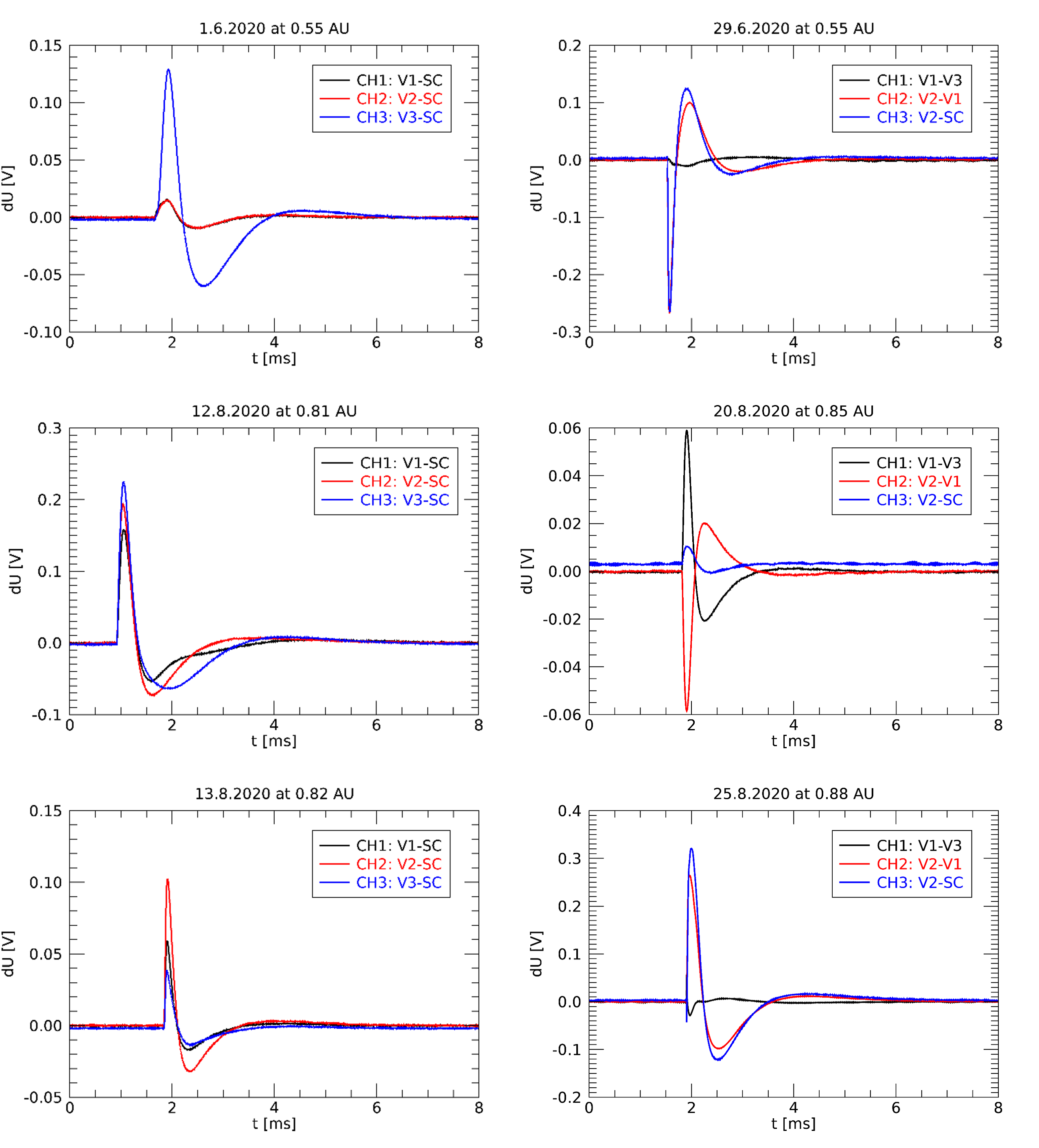}
   \caption{Six examples of TDS snapshots showing impulsive signals interpreted as due to dust impacts onto the spacecraft. Signals recorded on the three different channels are represented in different colors. The left column shows signals recorded in monopole mode (SE1 mode) whereas the right column shows signals recorded in dipole on the two first channels and monopole on the third one (XLD1 mode).}
              \label{fig_wf_examples}%
    \end{figure*}

Figure \ref{fig_wf_examples} shows examples of impulsive signals recorded by TDS that we interpret as due to dust impacts. The examples shown here were selected from the so-called triggered snapshots which were flagged by the on-board detection algorithm as dust probable impacts. See section \ref{sec-impactrate} for the description of the algorithm.

The left column shows events recorded in SE1 mode (for which TDS samples three monopoles V1, V2 and V3) whereas the right column shows events recorded in XLD1 mode (two channels measure dipole voltages V1-V3 and V2-V1, and the third channel is linked to the monopole V2). Each channel (CH1, CH2, and CH3) represents voltage difference measured between individual antennas (V1, V2, and V3) for dipoles or between one antenna and the spacecraft body (SC) for monopoles. It should be noted that presented data are not corrected for transfer function (the existence of “overshoots” in these data is probably artificial). Examples shown on the lower panels of left column are very typical of signals detected at the terminal of a monopole antenna by electron collection, similar to the one detected by STEREO/WAVES for instance.

\subsection{Voltage pulses and their link to mass and velocity of impacting dust grains}\label{voltage_model}

Before discussing these pictures, let us remind the mechanism through which voltage pulses are thought to be produced. First, a dust grain impacts the spacecraft body and expels from it some material, part of which is ionized. The amount of free electric charge $Q>0$ in the (overall neutral) cloud of expelled material is a function of the mass $m$ of the impacting grain and of the relative velocity $v$ of the impactor with respect to the target. The measurement of $Q$ after an impact in a dedicated collector, together with an independent measurement of $v$ in order to deduce the mass $m$ of the impactor, is the general operating principle of dust impact ionisation detectors \citep{Auer_chapter_2001}. Experiments show that the function $Q(m,v)$ follows the general scaling
\begin{equation}
Q(m,v) \simeq A m v^{\alpha}
\label{Q_mv}
\end{equation}
where $m$ is expressed in kg and $v$ in km.s$^{-1}$. Since the parameters $A$ and $\alpha$ quite strongly depends on the impacted material \citep{Collette_etal_2014}, it is of course preferable for the charge yield of the collector material as a function of $m$ and $v$ to be measured on ground. In the absence, for the moment, of such measurements for Solar Orbiter's surface material, we will have to use approximate values for $A$ and $\alpha$. \cite{Dietzel_etal_1973} or \cite{McBride_McDonnell_1999}) suggest the use of $A\sim1$ and $\alpha\sim3.5$, which is a rather high charge yield -- we shall discuss this point when evaluating the size of the impactors in the last section of this article.

In the absence of a dedicated and well calibrated collector, but in the presence of electric antennas operated in monopole mode, one can still quite reliably deduce the amount of charge $Q$ released during an impact, thanks to the different dynamics of the electrons and the heavier positive charges in the expelled cloud of ionized matter. The dynamics of the light electrons quickly decouple from the one of the heavier positively charged matter \citep{Pantellini_etal_CloudExpansion_2012}. The electrons are quickly collected by (or repelled away from) the spacecraft, letting positive charges unscreened in the vicinity of the spacecraft. For a positively charged spacecraft, it can be shown that the combination of the effect of quick electron collection and ions getting repelled away from the spacecraft will produce a maximal change in the spacecraft ground potential equal, to a good approximation, to $\delta \varphi_{sc} \simeq - Q/C_{sc}$ (here $C_{sc}$ is the spacecraft's body capacitance).

In monopole mode the signal recorded is $V(t) = \Gamma \left( \varphi_{ant}(t) - \varphi_{sc}(t) \right)$, where $\varphi_{ant}$ is the monopole antenna potential and $\Gamma$ a transfer function accounting for the (mostly capacitive) coupling between the antenna and the spacecraft body: $\Gamma = C_{ant} / (C_{ant} + C_{stray})$ (in this formula $C_{ant}$ is the antenna's capacitance and $C_{stray}$ accounts for the capacitive coupling through the preamplifier, the mechanical base of antenna and various wires). If one assumes that $\varphi_{ant}$ is roughly constant during the process, then the charge $Q$ produced by impact ionisation is quite simply linked to the peak of the voltage pulse measured in monopole mode by
\begin{equation}
Q(m,v) \simeq \frac{C_{sc} V_{peak}}{\Gamma}.
\label{Q_dV_monopole}
\end{equation}
The use of formulas (\ref{Q_mv}) and (\ref{Q_dV_monopole}) therefore makes it possible to link the properties of the impacting grain to those of the measured voltage signal.

We can see, in the light of these explanations, why monopole measurements are favourable to dust detection. The main changes induced by the impact occurs in the spacecraft ground potential, while antennas potentials stay roughly constant. Dipole measurements, which measure the variation of an antenna's potential relative to another antenna, are therefore quite insensitive to this process. Still, signals are quite frequently observed in dipole mode, as can be seen on the right panels of Fig.\ref{fig_wf_examples}.

For a signal to be observed in dipole mode, it must produce a signal quite stronger on a particular antenna than on the two others. An example of such a signal recorded on monopole mode can be seen on the top left panel of Fig.\ref{fig_wf_examples}, with a peak amplitude on monopole V3 much larger than on V1 and V2. An interpretation for these signals is that the impact location may be close to a particular antenna, the potential of which would in turn undergo a much stronger variation under electrostatic influence from the positively charged cloud than the other monopoles \citep{NMV_etal_GRL2014}. The derivation of the charge $Q$ from dipole measurements is therefore more complicated, since the amplitude of the voltage pulse not only depends on $Q$ but also on the position of the impact with respect to the monopoles. An order of magnitude of such a signal is $V_{peak} \sim \Gamma Q/(4\pi\epsilon_0 L_{ant})$ (assuming only one arm of the dipole sees the whole unscreened charge Q), so that the charge in the cloud can be linked to the peak voltage in dipole mode by
\begin{equation}
Q(m,v) \sim \frac{4\pi\epsilon_0 L_{ant} V_{peak, dipole}}{\Gamma},
\label{Q_dV_dipole}
\end{equation}
where $L_{ant}$ is the length of an antenna. Importantly, it should be noted that, unlike the signal observed in monopole mode (for the at least two monopoles showing similar peak voltages), which is quite accurately linked to the released charge $Q$ by eq.(\ref{Q_dV_monopole}), eq.(\ref{Q_dV_dipole}) only provides a rough order of magnitude, since the voltage produced will in fact be very dependent on the location of the impact. An impact occurring at equidistance from two arms of the dipole for instance, would produce a very small signal in dipole mode, even for an important release of charge, whereas an impact cloud expanding in the close vicinity of a particular dipole arm could produce a signal quite stronger than the estimation above.

On the right column of figure 1 (XLD1 mode) are shown a few examples of events where the signal is mainly registered by a single antenna - hence not mainly produced by the variation of the spacecraft potential, but rather by electrostatic influence on a particular antenna. The impact must be close to antenna V2 on cases shown on top and bottom right panels (that show identical pulses in channels CH2 and CH3) and to antenna V1 in the middle right panel (inverted pulses in channels CH1 and CH2).

\subsection{Parameters for Solar Orbiter as a dust detector}
\label{sec-orbiter-detector}

Now that we have detailed the main principles through which we interpret voltage pulses after a dust impact, we describe in this part their application to the specific case of Solar Orbiter.

Solar Orbiter orbits the Sun along a series of roughly elliptical orbits, with a minimum perihelion of 0.28 AU and a maximum inclination with respect to the ecliptic above $30^\circ$. A description of the mission and its science objectives can be found in \cite{Solo_science_2020}.

\begin{figure}
   \centering
   \includegraphics[width=\hsize]{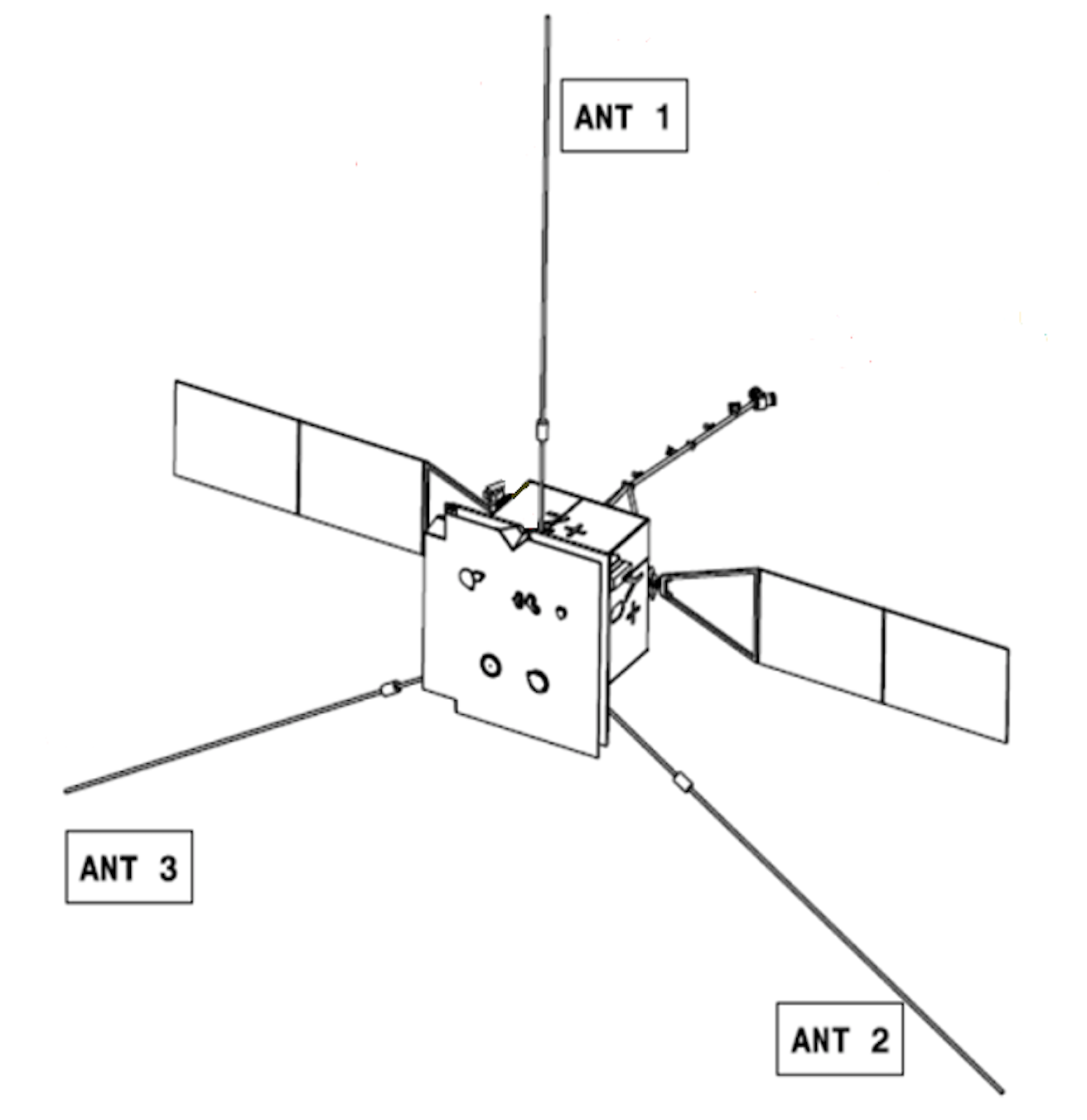}
   \caption{Geometrical configuration of RPW's electric sensors (labelled ANT1, 2 and 3), heat shield and deployed solar panels with respect to the spacecraft body.}
              \label{fig_solo_ants}%
    \end{figure}

RPW's electric sensors are three stacer monopoles of length $L=6.5$ m and radius $a = 0.015$ m, mounted on 1 m rigid booms to separate them from the spacecraft body, electrically biased in order to reduce variations of their potential with respect to the plasma at low frequencies \citep{Solo_RPW_2020}. They are in the same plane and make angles of roughly $120^\circ$ with each other. The disposition of the antennas with respect to the spacecraft body is shown on Fig.\ref{fig_solo_ants}. Let us note, in relation with the previous discussion on the generation of electric pulses after a dust impact, that the three monopoles are mounted on opposite sides of the spacecraft quite far away from each other (similarly to the case of spacecraft like WIND, MMS or Parker Solar Probe, but unlike the cases of Voyager, Cassini or STEREO), which implies that the effect of electrostatic influence can be very different from an antenna to the others, and explains why signals are frequently observed in dipole mode.

The capacitance associated to each monopole base has been measured on ground. Numbers found are $C_b = 76.3 \pm 4$ pF for the monopole V1, $C_b = 78.9 \pm 3$ pF for V2 and $C_b = 76.2 \pm 2.7$ pF for V3. The stray capacitance is to a good approximation the sum of this base capacitance and of the preamplifier capacitance $C_{amp}$, which, when both preamplifiers are ON, is $C_{amp} = 29$ pF. Since the base capacitances are equal within measurement uncertainties, we shall use for the three monopoles the same value of the stray capacitance, $C_{stray} \simeq 77 + 29 \simeq 108$ pF.

Assuming the monopoles are in vacuum and considering the spacecraft as an infinite ground plane, one finds the capacitance of a monopole $C_{ant} = 2\pi\epsilon_0 L / (\ln{L/a}-1) \simeq 71$ pF. The presence of solar wind's plasma, however, at the frequencies we are interested in (smaller than the local plasma frequency), decrease the capacitance to values $C_{ant} \sim 2\pi\epsilon_0 L / (\ln{L_D/a})$ where $L_D$ is the local Debye length \citep{MeyerPerche_1989}. With the Debye length varying in a range $\sim 3-10$ m along the spacecraft's orbit, one obtains $C_{ant} \sim 55-70$ pF (the capacitance increasing when going closer to the Sun). Using these values, the attenuation factor can be evaluated as $\Gamma = C_{ant} / (C_{ant} + C_{stray}) \simeq 0.34-0.39$.

The isopotential surface of the spacecraft’s body has been evaluated from the Solar Orbiter numerical model. This value can be estimated from 25.11 m$^2$, taking into account only the surface of the 5 satellite walls plus the heat shield, up to 31.43 m$^2$ when we consider the envelope including the overall spacecraft.. The solar panels (6 panels of $2.1\text{ m}\times 1.2$ m) are, at least their backside, isopotential to the spacecraft body. The additional surface is then $S_{panels} = 15.12$ m$^2$ (taking only one side into account). Therefore we can evaluate the area of the interface between the plasma and the isopotential parts of the spacecraft to be $S_{sc} \simeq S_{body}+S_{panels} \simeq 43.4 \pm 3.2 \text{ m}^2$. The capacitance of a conductor of such a complex shape is difficult to estimate. An order of magnitude estimation is given by the vacuum capacitance of the sphere of equivalent surface $C_{sc} \sim \epsilon_0 \sqrt{4\pi S_{sc}} \sim 210$ pF. This number is an underestimation that does not take into account sheath effects at the surface of the satellite between other effects, and we shall, in this paper use a value $C_{sc} = 250$ pF when this number is needed. A more appropriate determination of $C_{sc}$ could be made using a numerical model for Solar Orbiter and its interaction with the surrounding plasma, a work that will be undertaken in the future.

The surface $S_{sc}$ discussed above would approximately correspond to the dust collecting area if the dust population velocity distribution was isotropic in the frame of the spacecraft. As we will see in the following, it is probably not the case for the majority of the dust observed by Solar Orbiter, the velocity of which is mostly directed toward the heat shield. Therefore the dust collecting area to consider is strongly reduced compared $S_{sc}$, and is rather of the order of the heat shield's surface $S_{col} \simeq 2.5\text{ m}\times 3.1\text{ m} \simeq 8$ m$^2$.

Finally let us briefly discuss the relaxation time of perturbations of Solar Orbiter's floating potential. Linear theory gives $\tau_{sc} \simeq C_{sc} T_{ph} / I_e$, where $I_e = en_ev_eS_{sc}$ is the electron current onto the spacecraft isopotential surface ($e$ is the electron charge, $n_e$ the local electron density and $v_e = \sqrt{kT_e/(2\pi m_e)}$, with $T_e$ the local electron temperature and $m_e$ the electron mass). $T_{ph} \sim 3$ V is the temperature of the photoelectrons ejected from the spacecraft body expressed in electric potential unit. The assumption underlying this formula is that photoelectron current from the spacecraft is dominating over solar wind's electron current, the spacecraft potential being as a consequence positive (assumption justified pretty much all along Solar Orbiter's trajectory, putting apart short periods in the shadow of Venus). Assuming typical variation of electron parameters in the solar wind (see e.g. \cite{Issautier_etal_1998}) one obtains for the relaxation time $\tau_{sc} \sim 40$ $\mu$s at 1 AU and $\sim 10$ $\mu$s at 0.5 AU. These numbers have their importance in that the estimation of the peak amplitude of the pulses observed in monopole mode $V_{peak}\sim-Q/C_{sc}$ is strictly valid only in the case were the rise time of the pulse $\tau_{rise}$ (controlled by the positive charges dynamics in the vicinity of the spacecraft) isn't large compared to the relaxation time $\tau_{sc}$ of spacecraft electric potential perturbations. In the opposite case where $\tau_{sc}$ is small compared to $\tau_{rise}$ the amplitude of the signal is reduced by a factor of the order of $\tau_{sc}/\tau_{rise} \ll 1$ \citep{Zaslavsky_2015}. This effect will not be taken into account in this article. A more precise study of the waveforms -- which requires a very careful calibration work -- will be the subject of forthcoming studies, and will among other things make it possible to quantify this effect. 

\section{Statistical analysis of dust impacts}
\label{sec-stat-analysis}
\subsection{Impact rate and estimation of the impactors radial velocity}\label{sec-impactrate}

In this section we present the results of the analysis of the impacts voltage pulses recorded along Solar Orbiter's orbit from April 20th, 2020 to February 27th, 2021. For this purpose, we shall mainly use the data provided by the on-board analysis of TDS samples through an algorithm that flags a snapshot as being produced by a dust impact if it contains a signal impulsive enough. This dust detection algorithm has been working with constant settings from April 20th 2020, hence the date at which we start our analysis.

The detection algorithm described in detail in \citep{Solo_TDS_2021} works as follows: the instrument takes one waveform snapshot of 16384 samples every second. Each snapshot exceeding a minimum amplitude threshold is processed by the TDS on-board software to calculate the maximum and median amplitude and calculate the Fourier spectrum from the snapshot. From this spectrum the software determines the frequency corresponding to the largest spectral peak and the FWHM (full width at half maximum) bandwidth of this peak. The algorithm than classifies the observed snapshots based on comparing the ratio $R$ between the maximum and median absolute value in the snapshot and the spectral bandwidth to predefined thresholds. Specifically, snapshots with large $R$ and large spectral bandwidth are identified as dust impacts. This way, the algorithm allows to identify even relatively small amplitude dust spikes.

The outcome of this detection is then used to select the ``best'' wave and dust snapshots to be transmitted to the ground and also to build statistical data products. The key data product used here is the number of detected dust impacts in a 16 second interval which is transmitted in short statistical data packet every 16 seconds. Due to the fact that the detection algorithm only examines one snapshot of 62 ms every second, the reported dust counts are much lower than the actual number of dust impacts, but the number of detected dust impacts can be considered directly proportional to the actual number of dust impacts.

Since some impulsive signals may be erroneously taken for dust (e.g. solitary waves, \citep{Vaverka_etal_2018} and various spacecraft generated effects), the dataset has been cleaned by removing all periods of fast sampling at 524 kHz, all measurements influenced by active sweeps performed by the BIAS subsystem of RPW and several days (in particular during commissioning) where TDS was configured to non-standard operation modes. We also removed the Venus flyby interval on December 27th, 2020 when TDS detected numerous solitary waves and counted them as dust impacts.

We considered this TDS dust data product on a timescale of 1h, and computed the impact rate for each hour by dividing the number of snapshots flagged as dust by the total number of snapshots recorded during this hour multiplied by the duration of one snapshot ($\Delta t = 62$ ms): impact rate $R = N_{impact}/(N_{snapshots}\Delta t)$.

\begin{figure*}
   \centering
   \includegraphics[scale=0.6]{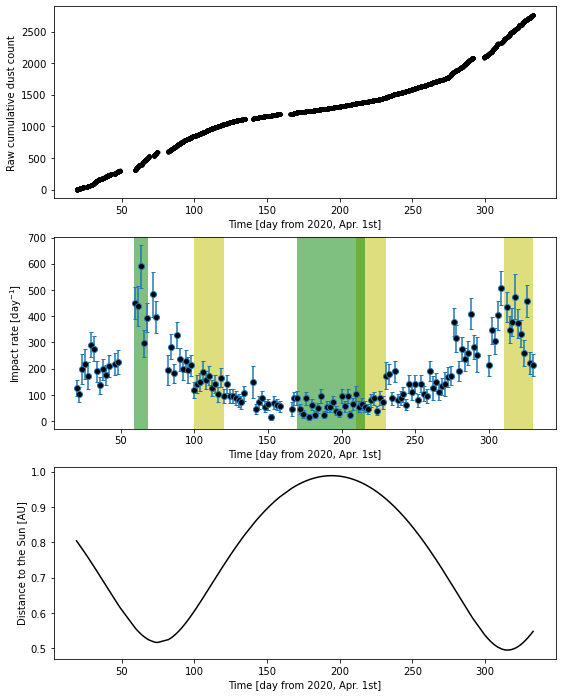}
   \caption{Upper panel: raw cumulative dust count as a function of time since April, 1st, 2020. By raw, it is meant that it is the integer number of impact provided by the detection algorithm, uncorrected for the instrument's duty cycle.
   Middle panel: Impact rate as a function of time since April, 1st, 2020. Each point correspond to a 2 days time interval. The error bars are computed from $\sqrt{N} / \Delta t$, where $\Delta t$ is the time covered by TDS measurements on the 2 days considered, and $N$ the number of impacts during these 2 days. The yellow areas correspond to time periods on which the number of occurrence distributions of Fig.\ref{fig_poisson_distributions} have been computed. The green areas correspond to times periods on which the amplitude distributions shown on the right panel of Fig.\ref{fig_amp_density} have been computed.
   Lower panel: Distance from the spacecraft to the Sun in astronomical units, as a function of time since April, 1st, 2020.}
              \label{fig_impact_rates}%
    \end{figure*}

Fig.\ref{fig_impact_rates} shows the evolution of the impact rates with time. The upper panel shows the raw (i.e. not corrected for the duty cycle) cumulative impact number, showing that the total number of impact detected by the algorithm during the period of our study is 2758. One can also notice several small data gaps corresponding to periods during which the instrument is switched off.

The middle panel shows the impact rate as a function of time, showing an increase of the flux with decreasing distance to the Sun, a general behaviour in agreement with remote and in-situ measurements from Helios \citep{Leinert_etal_1981} or Parker Solar Probe \citep{PSP_Szalay_2020, PSP_Page_2020}.

\begin{figure*}
   \centering
   \includegraphics[scale=0.45]{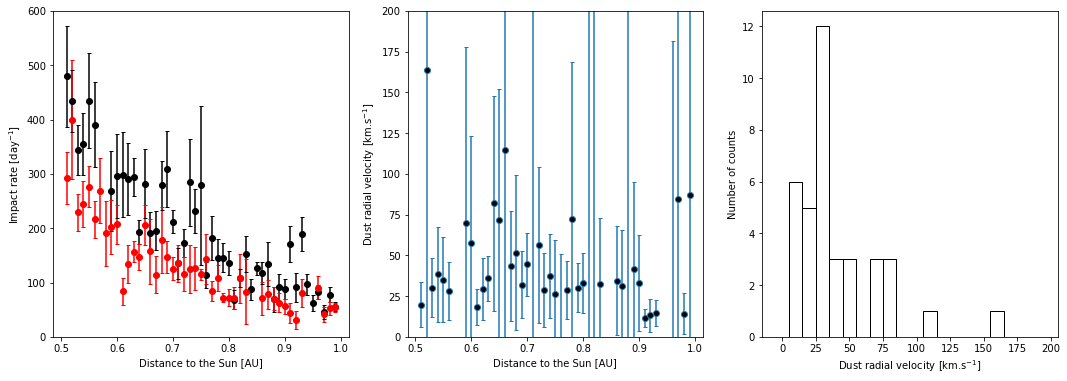}
   \caption{Left panel: impact rate as a function of radial distance. The black points show fluxes recorded when the spacecraft is moving toward the Sun, the red points when moving outward from the Sun. Each point is computed by averaging the impact rate data on 50 distance intervals linearly spaced between 0.5 and 1 AU. Error bars show one standard deviation error on the computation of the mean. Middle panel : radial component of the dust velocity $V_{r, dust}$ computed from eq.(\ref{eq_dust_velocity}). Error bars are computed by propagating errors on the impact rates shown on the left panel. Right panel : histogram of the obtained values of $V_{r, dust}$.}
              \label{fig_impact_rates_distance}%
    \end{figure*}

Fig.\ref{fig_impact_rates_distance} illustrates the evolution of the impact rate with distance from the Sun. On the left panel of this figure, the impact rates measured when the spacecraft is going toward the Sun (spacecraft radial velocity $v_{r, sc} < 0$) are separated from the ones measured when the spacecraft is going outward ($v_{r, sc} > 0$). It can be seen that the impact rate is in average slightly larger when $v_{r, sc} < 0$ than when $v_{r, sc} > 0$. This is consistent with the dust population measured having a mean velocity directed outward from the Sun. 

One can use this difference in the impact rates to obtain a first estimation of the radial velocity $v_{r, dust}$ of the dust population: assuming that the velocity and fluxes are function of the distance to the Sun only (neglecting all time variability) and neglecting -- for an order of magnitude estimation -- the effect of tangential velocities, it can easily be shown that 
\begin{equation}
v_{r, dust} \sim \frac{R_{in} + R_{out}}{R_{in} - R_{out}} \left\vert v_{r, sc} \right\vert 
\label{eq_dust_velocity}
\end{equation}
where $v_{r, sc}$ is the radial velocity of the spacecraft, $R_{in}$ the impact rate when the spacecraft is going toward the Sun and $R_{out}$ when going outward. The middle panel of Fig.\ref{fig_impact_rates_distance} shows the result of applying formula (\ref{eq_dust_velocity}) with the impact rates $R_{in}$ and $R_{out}$ shown on the left panel. The error bars are very large but the mean value obtained is reasonably constant with radial distance. The right panel shows a histogram of the values of $v_{r, dust}$ obtained, with a peak value at $v_{r, dust} = 30$ km.s$^{-1}$ and an averaged value around $50$ km.s$^{-1}$. The large errorbars, the variations of the measured fluxes and the use of a simple model let us only hope for an order of magnitude estimation of course, but this value is consistent with expectations for small particles on hyperbolic orbits, and to results from numerical simulations of particles dynamics discussed in the last section of this paper. 

\begin{figure*}
   \centering
   \includegraphics[scale=0.45]{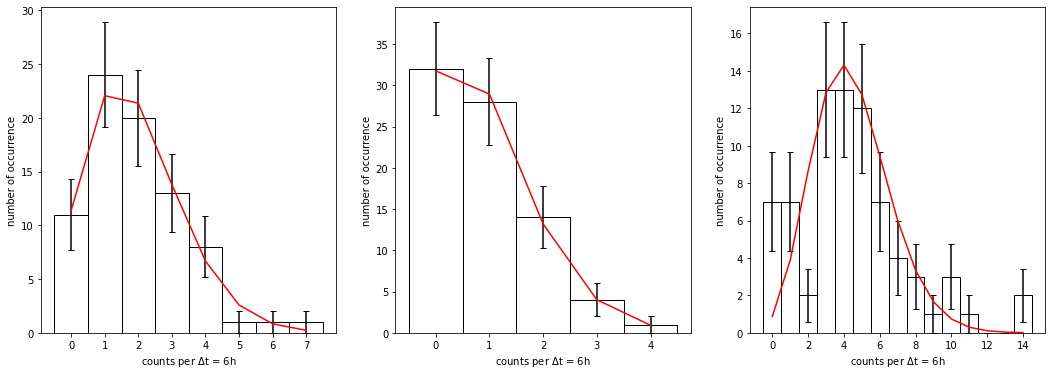}
   \caption{Distribution of number of impacts per time intervals of 6 hours, for three different periods of time corresponding to different mean impact rates. The red curve shows the Poisson distribution calculated from the mean impact rate observed in each of these periods of time. Errorbars on the histograms values are $\sqrt{N_{bin}}$, with $N_{bin}$ the number of counts in the histogram bin.}
              \label{fig_poisson_distributions}%
    \end{figure*}

Fig. \ref{fig_poisson_distributions} shows the distribution of the number of impacts recorded during time intervals of 6 hours, for three different values of the impact rate. The three intervals on which these number of occurrence distributions have been computed are highlighted in yellow on Fig.\ref{fig_impact_rates}. They correspond to distances to the Sun $\sim 0.7$ AU (left panel, mean impact rate $\sim 140$ day$^{-1}$), $\sim 1$ AU (middle panel, mean impact rate $\sim 50$ day$^{-1}$) and $\sim 0.5$ AU (right panel, mean impact rate $\sim 350$ day$^{-1}$). The comparison with Poisson distribution, over-plotted in red, shows a very good agreement consistent with the data being due to independent events at roughly constant rates, as expected for interplanetary dust impacts.

\subsection{Peak voltages distribution and power-law index of the impactor's cumulative mass flux}

In order to further characterize the population of dust grains impacting Solar Orbiter, we look at the distribution of voltages measured in monopole mode. To this purpose we look at the snapshots dataset, which does not include all of the dust impacts detected by the onboard algorithm on which the results of the previous section are based. We shall assume that the subset of triggered snapshots is a random sample from the ensemble of all the recorded dust impact signals, and therefore that both share the same statistical properties -- or since this can't be exactly true, we shall assume that selection bias are small and do not impact importantly the voltage amplitude statistics.

Measurements in monopole mode, as discussed in section 2, are required in order to properly (as unambiguously as possible) link the peak amplitude of the pulse to the charge generated by impact ionisation. Unfortunately they are, concerning monopoles V1 and V3, only active during a small fraction of the mission time, i.e. mainly during two periods, from May 30th to June 8th (185 dust snapshots telemetered) and from July 27th to August 13th (103 dust snapshots). Things are different for monopole V2, which is quite routinely operated with 934 dust snapshots telemetered from April 1st to November 1st of 2020. Hence the particular focus on monopole V2 in the following.

\begin{figure*}
   \centering
   \includegraphics[scale=0.45]{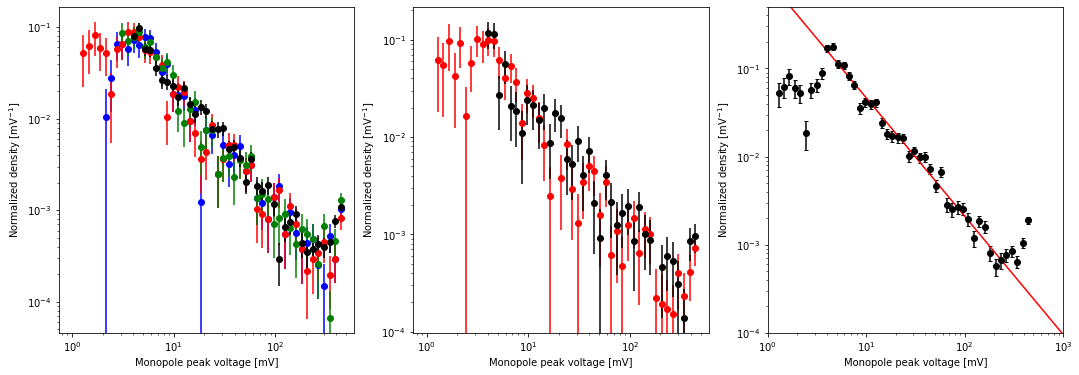}
   \caption{Normalized density (i.e. number of dust detected per unit voltage interval) of voltage peaks in monopole mode. The left panel shows the distribution of the voltage peaks on all monopoles from Apr. 1st to Nov. 30th, 2020. Different colors accounts for different monopoles: V1 (blue), V2/SE1 (red), V3 (green), V2/XLD1 (black). The middle panel shows the distribution of peak amplitudes on monopole V2 during two different periods (highlighted in green on fig.\ref{fig_impact_rates}). Red points correspond to the high impact rate period close to the perihelion (May 13th -- Jun 8th), black points to the low impact rate period around the aphelion (Sep 17th -- Nov 3nd). The right panel shows the result of the least square fitting of all the data recorded on monopole V2 from Apr. 1st to Nov. 30th, 2020. The slope of the red curve is $-1.34 \pm 0.07$. On all these figures, errorbars are computed from $\sqrt{N_{bin}}/(\Delta V_{bin}N_{tot})$, where $N_{bin}$ is the number of counts in the bin, $\Delta V_{bin}$ the bin width in mV and $N_{tot}$ the total number of events from which the histogram is computed.}
              \label{fig_amp_density}%
    \end{figure*}

Fig.\ref{fig_amp_density} shows the normalized distributions of peak voltage associated to dust impacts for different monopoles, and different time periods. We can see that the peak voltage distribution has a power-law behaviour. The left panel shows distributions computed on the whole period Apr. 1st - Nov. 31st, with different colors corresponding to peak voltages at the terminals of different monopoles. One can see that voltage distributions are similar on all monopoles - which is to be expected since in average all monopole should react in the same manner to dust impacts. The results of least-square fitting these histograms with power laws is presented in table \ref{table_fits}, where it can be seen that power-law exponents are, within uncertainties, equal from a monopole to another, around $-4/3$.

\begin{table*}
\caption{Power law indices of the peak voltage distributions. The uncertainties show 95$\%$ confidence interval on the linear regression coefficient.}             
\label{table_fits}      
\centering                          
\begin{tabular}{c c c c}        
\hline\hline                 
Monopole  &  Time interval & Number of events & Power-law index \\   
\hline\hline                        
   V1 (SE1 mode) & Apr. 1st -- Nov. 31st & 328 & $-1.34\pm0.14$    \\  
   V2 (SE1 mode) & Apr. 1st -- Nov. 31st & 328    & $-1.34\pm0.14$ \\
   V2 (XLD1 mode) & Apr. 1st -- Nov. 31st & 934     & $-1.37\pm0.10$ \\
   V3 (SE1 mode) & Apr. 1st -- Nov. 31st & 328    & $-1.36\pm0.11$ \\
\hline                                   
   V2 (SE1 mode) & May 30th -- Jun. 8th  (Perihelion) & 185  & $-1.37 \pm 0.19$ \\
   V2 (XLD1 mode) & Sep. 17th -- Nov. 2nd (Aphelion) & 161    & $-1.20 \pm 0.17$ \\
\hline  
   V2 (SE1 and XLD1 modes) & Apr. 1st -- Nov. 31st & 1262    & $-1.34 \pm 0.07$ \\

\hline \hline                                
\end{tabular}
\end{table*}

The middle panel of Fig.\ref{fig_amp_density} compares the peak amplitude recorded on monopole V2 at different distances from the Sun. The limited number of impacts on which such a comparison can be made imply quite important uncertainties on the values of the distribution. The two last lines of table \ref{table_fits} show the results of linear fitting for these distributions, showing a slope a bit steeper at the perihelion than at the aphelion. This difference being within the uncertainties, it is hard to conclude on this result; one should wait for more statistics to see if this trend is confirmed.

The right panel of Fig.\ref{fig_amp_density} shows the distributions of all impacts recorded on monopole V2 on the whole time of our voltage amplitude analysis, i.e. from Apr. 1st to Nov. 31st, 2020, and the corresponding power-law least-square fit, with index $-1.34 \pm 0.07$.

From these observations it seems reasonable to approximate the rate of observation of signals having peak voltages between $V_{peak}$ and $V_{peak}+dV_{peak}$ by $dR = g(V_{peak}) dV_{peak}$, with 
\begin{equation}
g(V_{peak}) = g(V_{0}) \left( \frac{V_{peak}}{V_0}\right)^{-a}
\label{f_mass}
\end{equation}
where $a \simeq 1.34$ and $V_0$ an arbitrary voltage in the range where the power-law behaviour applies.
 
An interesting result, of course, would consist in the derivation from these data of an information on the mass distribution of the impacting dust particles. We have seen in section \ref{voltage_model} that the released charge $Q$, and hence the peak voltage $V_{peak}$ is linked to the mass, but also to the velocity of the impacting particle, and we do not have an independent measurement of the latter for each of the impacts. Therefore one can only make inferences on the mass distribution by assuming the existence of a relationship  $v(m)$ between the mass of the particle and its velocity with respect to the spacecraft. If such a relationship exists, then the function $V_{peak}(m, v)$ becomes a function of $m$ only and -- under the assumption that two different values of $m$ cannot produce a peak voltage of the same amplitude, i.e. that the function $V_{peak}(m)$ is bijective on the observed mass interval -- the mass distribution of the impactors $f(m)$ can be directly linked to the  measured distribution $g(V_{peak})$ by
\begin{equation}
f(m) = g(V_{peak})\left\vert\frac{dV_{peak}}{dm} \right\vert.
\label{f_mass}
\end{equation}

For a first order estimation, one could assume that the impact velocity is independent of the mass on the given mass range. Then, according eqs.(\ref{Q_mv}) and (\ref{Q_dV_monopole}),
\begin{equation}
V_{peak} \simeq \frac{\Gamma}{C_{sc}}A m v^\alpha
\label{V_peak}
\end{equation}
is a linear function of $m$, and  the mass distribution trivially has the same power-law shape than the distribution of voltage peaks. The cumulative mass flux of particles of mass larger than $m$ onto the spacecraft (defined as $F(m) = \int_m^{+\infty} f(m')dm'$) would then be given by
\begin{equation}
F(m) = F(m_0)\left( \frac{m}{m_0}\right)^{-\delta}
\label{F_mass}
\end{equation}
where $\delta = a-1 \simeq 0.34$ and $F(m_0)$ is the cumulative flux of particles of mass larger than $m_0$ (that may depend on the distance to the Sun).

Let us note that this estimation of the power-law index $\delta$ of the cumulative mass flux in the observed mass range must likely be an underestimation, since if the velocity is an increasing function of $m$ (which is probably the case in the observed mass range, see next section), $f(m)$ will decrease with a steeper slope than $g(V_{peak})$. This can easily be seen from eq.(\ref{f_mass}), taking for instance $V_{peak} \propto m^{1+\varepsilon}$ with $\varepsilon>0$. One would then obtain for $f(m)$ a power law index $a + (a-1)\varepsilon$ which is always larger than $a$ (if $a>1$ which is the case here). Therefore, even without any precise knowledge of the function $v(m)$, but under the assumption that this is an increasing function of $m$ in the observed mass interval, it is possible to derive  from these observations of peak voltages a lower bound for the power law index of the cumulative mass flux $\delta \gtrsim 0.34$.

To conclude this part let us note that a more detailed derivation of the mass distribution of the particles could be obtained from these measurements by computing the function $v(m)$ from numerical simulations, with proper assumptions on initial conditions and dust $\beta$ parameter. Since this function may depend on the distance from the Sun (which may explain the possible change of the power law index of the voltage distributions from perihelion to aphelion), this study may also require some time to accumulate more statistics and be able to construct not too noisy distributions of voltages at different distances from the Sun. Such a work is beyond the scope of this first results paper, but it offers an interesting perspective for a future study.

\section{Estimation of the $\beta$-meteoroids flux and comparison to models and results from other missions}

We now compare the observed impact rates to a dust flux model that bases on a number of assumptions. The existence of dust in the inner solar system can be inferred from the brightness of the Zodiacal light and the F-corona which show that dust in the approximate 1 to 100 micrometer size range forms a cloud with overall cylindrical symmetry about an axis through the center of the Sun, perpendicular to the ecliptic (cf. \cite{Mann_etal_2014}). The size distribution at 1 AU is estimated from a number of different in-situ observations and described in the interplanetary dust flux model \citep{Gruen_etal_1985}. The majority of dust grains form by fragmentation from comets, asteroids and their fragment grains. The larger grains are in bound orbit about the sun; as a result of the Poynting – Robertson effect they lose orbital energy and angular momentum and fall into the Sun after time scales of the order of $10^5$ years. However, the majority of the Zodiacal dust is within shorter time fragmented by collision with other dust grains, the smaller fragments leave the inner solar system and collision production from larger grains is needed to replenish the dust cloud \citep{Mann_etal_SSR2004}. The dust with sizes smaller than roughly a micrometer experiences a larger radiation pressure force which is directed outward. If the radiation pressure to gravity ratio, often denoted as $\beta$ is sufficiently high, the dust can be ejected outward. Those grains in hyperbolic orbits are often denoted as $\beta$-meteoroids those in in bound orbits as $\alpha$-meteoroids \citep{Zook_Berg_1975}. Assuming that the larger dust grain that is fragmented moves in a circular orbit, its fragment attains a hyperbolic orbit if the radiation pressure to gravity ratio $\beta$ exceeds 0.5. Based on light scattering models for different assumed dust compositions \citep{Wilck_Mann_1996}, this is the case for dust with sizes less than few 100 nm. 

\subsection{Mass of the impactors}

The mass of the impactors can be estimated from the voltage observed, using eq.(\ref{V_peak}) and spacecraft parameters from section \ref{sec-orbiter-detector}. For this one needs an estimation of the velocity of the impactors and of the charge yield of the impacted material. For the particles velocity, we could have used the order of magnitude obtained from the observations of impact rates in direction forward and backward with respect to the Sun. But, as pointed out in section \ref{sec-impactrate} this is only a rough estimation, associated to large uncertainties. Therefore we chose to rely in this section on estimations of dust particles velocities from numerical simulations of dust dynamics in the interplanetary medium. The simulation we use takes into account gravitational and radiation pressure forces, but not the electromagnetic force (that should not be dominant for particles of sizes $\gtrsim 40$ nm). Initial conditions and values of $\beta$ are chosen consistently with observations of dust from asteroidal origin.

\begin{figure*}
   \centering
   \includegraphics[scale=0.5]{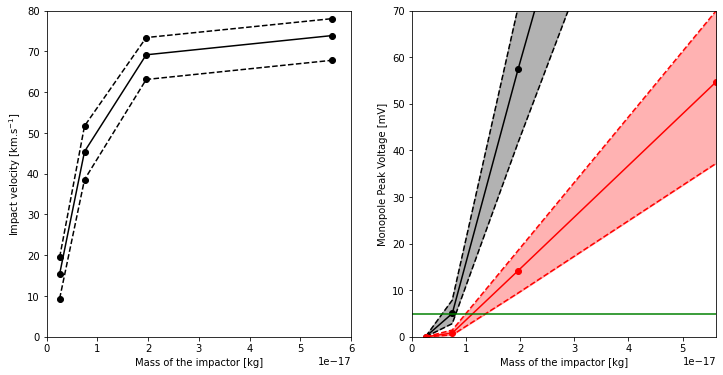}
   \caption{Left panel : relative velocity with respect to Solar Orbiter as a function of particle mass. The solid line shows the average along the spacecraft trajectory, the upper dahsed line shows the impact velocity at the perihelion and the lower dashed line the impact velocity at the aphelion. Right panel : expected peak voltage as a function of the mass of the impactor, calculated using values of impact velocities shown on the left panel. The solid line shows the peak voltage for the averaged velocity (solid line from left panel) and dashed lines for the perihelion and apehlion impact velocities. The black curves assume a high charge yield with parameters $A = 0.7$ and $\alpha=3.5$ and the red curves a lower charge yield with parameters $A = 2.5\times10^{-3}$ and $\alpha=4.5$. The points show the values of the velocity/peak voltage obtained from 4 numerical simulations, corresponding to size ranges $40-75$ nm, $75-100$ nm, $100-140$ nm and $140-200$ nm. The points are placed at the middle of the mass interval for each of the 4 simulated size ranges. The horizontal green line shows the detection threshold $\sim 5$ mV.}
              \label{fig_velocity_and_peak_voltage}%
    \end{figure*}

The charge yield of the impacted material is another unknown of our study, which of course plays an important role in determining the mass of the impactors. Fig.\ref{fig_velocity_and_peak_voltage} shows the impact velocities from the numerical simulations and the expected peak voltage for two examples of charge yield. The dark area on the right panel shows the region of expected voltage peaks as a function of the mass of the impactor for parameters $A = 0.7$ and $\alpha=3.5$ from \cite{McBride_McDonnell_1999}. The red area is obtained for parameters $A = 2.5\times10^{-3}$ and $\alpha=4.5$, a quite lower charge yield, corresponding to materials like germanium-coated Kapton, or solar cells and MLI (multi-layer thermal insulation) from STEREO's spacecraft, the charge yield of which were measured on ground and can be found in table 1 of \cite{Collette_etal_2014}.

Fig. \ref{fig_velocity_and_peak_voltage} shows that particles in the size range $40-75$ nm, regardless of the charge yield parameters used, are too small and not fast enough to produce measurable signals. Grains with sizes $75-100$ nm should produce (for the larger of them) signals above the detection threshold in the high charge yield case, but not in the low charge yield one; grains with sizes $\gtrsim 100$ nm finally, should produce measurable signals regardless of the precise charge yield. An estimation of the mass of the smallest particles detected (we consider a threshold voltage of 5 mV) from this figure gives $m \simeq 7.4\times 10^{-18}$ kg (high charge yield) and $m \simeq 1.2 \times 10^{-17}$ kg (low charge yield), corresponding to sizes (assuming, as in the whole of this paper, a mass density $\rho=2.5$ g.cm$^{-3}$ and grains of spherical shapes for size-mass conversions) $r \simeq 89$ nm and $r \simeq 103$ nm respectively -- one can quite confidently affirm that the smallest grains detected have sizes around 100 nm.

The curves for different charge yields diverge when masses are increased, and for signals of amplitude e.g. $30$ mV, one have sizes $r \simeq 107$ nm (high yield) and $r \simeq 150$ nm (low yield), a larger mismatch of course, showing the necessity of ground measurement if one wants to reach a better mass calibration on the whole voltage interval. That being said, the power-law decrease of the peak voltages distribution shown in the previous section implies that the fluxes are dominated by impact from small grains, so that the determination of the mass of impactors producing high amplitudes peaks is less critical for our study.

This discussion of particles masses is based on the measurement of the voltage pulses in monopole mode, which, as was discussed previously, are the most reliable when it comes to estimating the charge released by impact ionisation, and therefore the dust particles masses. However, the dust detection algorithm from which fluxes are computed works on a TDS channel that is operated in dipole mode most of the time -- when the instrument is on XLD1 mode. The discussion above, and the curve presented on Fig.\ref{fig_velocity_and_peak_voltage} stays mainly relevant to this case as long as the signal produced is proportional to the charge released $Q$. According to the estimation given by eq.(\ref{Q_dV_dipole}) this should be the case for dipole measurements when considering a large enough number of hits. Assuming $V_{dipole} = \Gamma Q/ 4\pi\epsilon_0 L_{ant}$, the smallest mass measured in dipole would be a factor $4\pi\epsilon_0 L_{ant} / C_{sc} \sim 3$ larger (and therefore the sizes a factor $\sim 1.4$ larger). But the precise factor is complicated to evaluate, and could be closer to 1 in average since observations of waveforms shows that differences in peak voltage from a monopole to another is often of the order of magnitude of the peak voltage itself. This provides a clear motivation for for developing a quantitative theoretical modeling of the generation of signals generation in dipole mode, at least on a statistical basis. 

\subsection{Flux of $\beta$-meteoroids, comparison to predictions and measurements from other spacecraft}

\begin{figure*}
   \centering
   \includegraphics[scale=0.6]{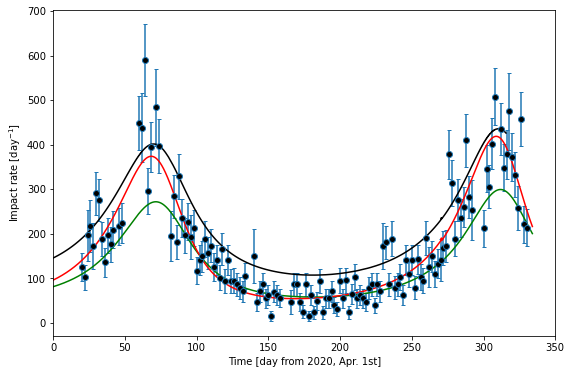}
   \caption{Dust impact rate onto the spacecraft - the dots show the same data as the one presented on fig.\ref{fig_impact_rates}. The green curve shows the expected impact rate for beta meteoroids having a constant outward velocity $v_\beta = 50$ km/s, and a flux density $F_{1AU} = 8\times 10^{-5}$ m$^{-2}$s$^{-1}$ at 1 AU. The red curve shows the flux from eq.(\ref{rate_th_beta_mass_thres}) with $F_{1AU} = 8\times 10^{-5}$ m$^{-2}$s$^{-1}$ and $\alpha\delta = 1.3$. The black curve shows the prediction from numerical simulation based on a model of production of dust by collisional fragmentation. The collecting area is assumed to be equal to the heat shield's surface, $S_{col} = 8$ m$^2$.}
              \label{fig_rate_compa_model}%
    \end{figure*}

Fig.\ref{fig_rate_compa_model} presents the impact rates averaged over 2 days, already presented on Fig.\ref{fig_impact_rates} of this article, together with impact rates computed for three models of the beta-meteoroid flux. The green line is the most simple model, with the impact rate given by
\begin{equation}
R = F_{1AU} S_{col} \left( \frac{r}{\text{1 AU}} \right)^{-2} \frac{v_{impact}}{v_{\beta}}
\label{rate_th_beta}
\end{equation}
where $S_{col} = 8$ m$^2$ is the collection area (taken equal to the heat shield surface, cf. section \ref{sec-orbiter-detector}), $v_{impact} = \left\vert\left\vert v_\beta - v_{sc} \right \vert \right\vert$ is the relative speed between the spacecraft and the $\beta$-meteoroids, the velocity of which is taken radial and constant $v_\beta = 50$ km.s$^{-1}$.  $F_{\beta, 1AU}$ is the flux of particles in the detection range, which is to a good approximation equal to the cumulative flux of particles larger than $\sim 100$ nm. The $1/r^2$ scaling of the $\beta$-meteoroids flux is a consequence of mass conservation (and of their production rate by fragmentation of larger particles being negligible at the distances from the Sun at which the spacecraft orbits). A fit of the data with eq.(\ref{rate_th_beta}) gives a value of the cumulative flux $F_{1AU} \simeq 8\times 10^{-5}$ m$^{-2}$s$^{-1}$.

This simple model provides, as can be seen on the figure, a fairly good agreement with the data, although it seems to systematically underestimate the high impact rate values around the perihelion. A reason for that can be seen from the left panel of Fig.\ref{fig_velocity_and_peak_voltage}: the mass of an impactor producing a given peak voltage will be smaller at the perihelion than at the aphelion (because the spacecraft's velocity with in the Sun's frame is larger at the perihelion that at the aphelion). Since smaller particles have higher fluxes, an increase in the rate of measurable signals is expected close to the Sun as compared to the previous simple model.

This effect of decrease of the dust particles mass measured when approaching the Sun can be quantified by assuming a cumulative mass flux varying in power law with an exponent $\delta$, as written in previous section's eq.(\ref{F_mass}). The impact rate is then found to be
\begin{equation}
R = F_{1AU} S_{col} \left( \frac{r}{\text{1 AU}} \right)^{-2} \frac{v_{impact}}{v_\beta} \left(\frac{v_{impact}}{v_{impact}(\text{1 AU})}\right)^{\alpha\delta}
\label{rate_th_beta_mass_thres}
\end{equation}
where $F_{1AU}$ is (as previously) the cumulative flux of particles above the detection threshold at 1 AU, and the factor $(v_{impact}/v_{impact}(\text{1 AU}))^{\alpha\delta}$ accounts for the variation of the mass of the detected impactors with the impact velocity. 

One can fit eq.(\ref{rate_th_beta_mass_thres}) to the data in order to obtain the product $\alpha\delta$. The resulting curve is shown in red on Fig.\ref{fig_rate_compa_model}. It shows a better agreement with data than the previous model, and fits quite well the high impact rates observed at the perihelion. The value obtained is $\alpha\delta = 1.3$, which, considering values of $\alpha = 3.5 - 4.5$, provides an estimation of the power-law index of the cumulative mass distribution of the impactors $\delta = 0.29 - 0.37$. This value -- obtained from a dataset including no voltage measurements but only impact counts per unit time -- is similar to the one derived in the previous section by fitting peak voltage distributions, indicating that the estimation of $\delta$ in RPW's detection range seems quite robust. Let us note finally that, assuming the measurement of $\delta$ from voltage distributions to be reliable enough, one could use this estimation of $\alpha\delta$ to independently estimate the power-law index $\alpha$ of the charge yield for Solar Orbiter material. We would then obtain $\alpha \simeq 1.3/0.34 \simeq 3.9$.

The black curve on Fig.\ref{fig_rate_compa_model}, finally, shows the prediction of impacts onto the spacecraft from a model of production of small dust grains by collisional fragmentation. This model assumes that the parent bodies move in Keplerian orbits within the  circumsolar dust disk. Their mass distribution is a modified version of the interplanetary dust flux model \citep{Gruen_etal_1985}. The size distribution of the collision  fragments are described based on models by \cite{Tielens_etal_1994} and \cite{Jones_etal_1996}. It describes the fragmentation and partial vaporization of a target and projectile composed of a certain dust material. The vaporized and fragmented mass of the target are proportional to the projectile mass and to the velocity -- and material-dependent coefficients. The mass distribution of fragments is of the form $m^{-0.76}$, with the largest fragment mass specified as some (collision-velocity dependent) fraction of the target mass. A brief description of the collision model is given in \cite{Mann_Czech_2005}. The derivation of dust fluxes is described in \cite{PSP_Mann_Czechowski_2021}. They are obtained from the dust trajectories under the influence of gravity and radiation pressure (since the Lorentz force does not have a strong influence for the considered dust sizes). The same trajectories were used to produce Fig.\ref{fig_velocity_and_peak_voltage} of this article. The black curve shows the prediction from this model of the number of particles comprised between 100 nm and 200 nm impacting a surface $S_{col}=8$ m$^2$ per day. The curve obtained from the model considers a constant minimal size of detected dust at $100$ nm all along the trajectory. It does not include the effect of variation of minimal mass detected discussed previously. It fits, without varying any free parameter, well the high impact rates at perihelion but quite overestimates the low fluxes period - a result that would tend to indicate that the size of the particles detected is closer to 100 nm at the perihelion and probably a bit larger at aphelion.

Some effects that can play a role on the derivation of the particle flux from the observed impact rates have not been taken into account in this first study. They include a better model for the collection surface and its possible variation with distance to the Sun (the direction of dust's velocity in the spacecraft frame varying along the trajectory), although given the mostly cubic shape of the spacecraft this effect is not expected to be very important. A possible difference in charge yield for an impact on the heat shield surface and one of the other 5 spacecraft walls could also have some importance.

The result that we obtained for the $\beta$-meteoroid flux is similar (although slightly higher) than the one derived using STEREO/WAVES by \cite{Zaslavsky_etal_2012}, of $F_\beta \simeq 1-6 \times 10^{-5}$ m$^{-2}$s$^{-1}$ at 1 AU. Similar also to the measurement with Solar Probe Plus FIELD instrument $F_\beta \simeq 3-7 \times 10^{-5}$ m$^{-2}$s$^{-1}$ interpolated at 1 AU \citep{PSP_Szalay_2020}. Similar, finally, to theoretical expectations, as shown in this paper.

To conclude, let us note that, if the presented data can be well described with a flux of $\beta$-meteoroids, it lacks, in comparison with observations from STEREO \citep{Zaslavsky_etal_2012, Belheouane_ISD_2012} or Wind \citep{Malaspina_etal_2014} an observed flux of interstellar dust. The latter studies indeed showed a noticeable component of the impact rate modulated along the solar apex direction, that is not observed with Solar Orbiter RPW. This lack of an apparent interstellar dust component in the data is puzzling. It could be explained by a deflection of the interstellar dust grains in the solar magnetic field, and the consequent depletion of their flux inside 1 AU -- which is expected for grains of small size \citep{Mann_ISD_2010}.
This is a point which deserves further studies.

\section{Conclusions}

   \begin{enumerate}
   \item The analysis of the first data from Solar Orbiter Radio and Plasma Wave instrument shows this instrument to be a quite reliable dust detector for dust grains in the size range $\gtrsim$ 100 nm. Fluxes of particles derived are consistent with previous observations in this size range and with theoretical prediction from models of dust production by collisional fragmentation.
   \item The analysis of the difference in impact rates when the spacecraft's velocity vector is directed in the sunward or anti-sunward direction is shown to be able to provide a direct measurement of the order of magnitude of dust grains radial velocities $v_{r, dust} \sim 50$ km.s$^{-1}$ consistent with theoretical predictions in the observed size range.
   \item The analysis of voltage distribution in monopole mode, and the analysis of the impact rates taking into account a variation of the mass of the smallest grains detected as a function of the impact velocity provide two independent methods for estimating the power law index $\delta$ of the cumulative mass flux of particles in our detection range. These two methods consistently provide a result $\delta \simeq 1/3$. 
    \item These first results are on overall very promising. Still, numerous works remain to be done, in particular a modeling of the signal generation mechanism in monopole mode that would including the effect of variation of the floating potential relaxation time with the local plasma parameters, but also models of signal generation in dipole mode. Such works, together with a more precise modeling of grain's velocity and additional statistics could make it possible to derive more information on the dust cumulative mass flux from the peak voltage distributions.
    \item Solar Orbiter will reach perihelion close to 0.3 AU in spring 2022. The impact rate should be $\sim$ 3 times (without taking the mass detection threshold effect) larger at this point than at the perihelion studied in this article. Solar Orbiter's orbit will also reach increasingly higher latitudes in the years to come, and will provide the first in-situ exploration of the dust cloud out of the ecliptic. These perspectives are very promising, and these first results show that RPW will have the capabilities to provide scientific return from these opportunities.
        \end{enumerate}

\begin{acknowledgements}
IM has been supported by the Research Council of Norway (grant no. 262941). JV was supported by the Czech Science Foundation under the project 20–13616Y. MM and YK are supported by the Swedish National Space Agency grant 20/136. AZ, IM and JV acknowledge discussions during the ISSI team on dust impacts at the International Space Science Institute in Bern, Switzerland.
\end{acknowledgements}

%
%

\bibliographystyle{aa}
\bibliography{so_dust-2col}

\end{document}